\documentclass[manuscript]{acmart}

\AtBeginDocument{%
  \providecommand\BibTeX{{%
    \normalfont B\kern-0.5em{\scshape i\kern-0.25em b}\kern-0.8em\TeX}}}

\setcopyright{acmcopyright}
\copyrightyear{2023}
\acmYear{2023}
\setcopyright{rightsretained}
\acmConference[RecSys '23]{Seventeenth ACM Conference on Recommender Systems}{September 18--22, 2023}{Singapore, Singapore}
\acmBooktitle{Seventeenth ACM Conference on Recommender Systems (RecSys '23), September 18--22, 2023, Singapore, Singapore}\acmDOI{10.1145/3604915.3608883}
\acmISBN{979-8-4007-0241-9/23/09}

\usepackage{subcaption}

\begin{document}

\title{Beyond Labels: Leveraging Deep Learning and LLMs for Content Metadata}


\author{Saurabh Agrawal}
\email{sagrawal@tubi.tv}
\orcid{0000-0003-2233-1604}

\author{John Trenkle}
\email{jtrenkle@tubi.tv}
\orcid{0009-0005-2238-2024}
\author{Jaya Kawale}
\email{jkawale@tubi.tv}
\orcid{0009-0005-5451-3308}
\affiliation{%
  \institution{Tubi}
  \streetaddress{315 Montgomery Street, Fl 16}
  \city{San Francisco}
  \state{California}
  \country{USA}
  \postcode{94104}
}


\renewcommand{\shortauthors}{Agrawal, et al.}


\begin{abstract}
Content metadata plays a very important role in movie recommender systems as it provides valuable information about various aspects of a movie such as genre, cast, plot synopsis, box office summary, etc. Analyzing the metadata can help understand the user preferences to generate personalized recommendations and item cold starting. In this talk, we will focus on one particular type of metadata - \textit{genre} labels. Genre labels associated with a movie or a TV series help categorize a collection of titles into different themes and correspondingly setting up the audience expectation. We present some of the challenges associated with using genre label information and propose a new way of examining the genre information that we call as the \textit{Genre Spectrum}. The Genre Spectrum helps capture the various nuanced genres in a title and our offline and online experiments corroborate the effectiveness of the approach. Furthermore, we also talk about applications of LLMs in augmenting content metadata which could eventually be used to achieve effective organization of recommendations in user's 2-D home-grid.
\end{abstract}




\maketitle
\section{Introduction}

Online video-on-demand streaming services, like Tubi, have become essential sources for consuming entertainment online. Offering personalized recommendations to users is crucial for helping them discover content that aligns with their preferences. Content metadata, including genre, director, cast, plot synopsis, and box office summary, provides valuable complementary information about different aspects of a title, greatly contributing to the personalization process. In this talk, our focus lies on genre labels, which represent broad categories such as "horror," "comedy," or "romance," and have demonstrated significant effectiveness in tailoring the user experience ~\cite{pirasteh2021personalized, duong2020genres}.


\textbf{Challenges:} Extracting valuable signals directly from genre labels presents several challenges. Firstly, the absence of a universally accepted definition of genre and the lack of consensus on genre categorizations from various different sources pose significant obstacles. While genres like romance, comedy, thriller, action, drama, and horror are widely recognized and prevalent, there are lesser-known genres such as sports, documentary, western, music, etc., some of which may not be universally acknowledged as genres. Consequently, genre labels can be noisy, inconsistent, and subjected to the personal judgments of annotators. Secondly, the coverage of less-common genres in movies tends to be limited, reducing the availability of data for these genres. This scarcity poses challenges for accurately capturing and representing the characteristics of these genres. Thirdly, certain genres, like "rom-com," are more of a combination of other genres. These blended genres further complicate the categorization process and introduce ambiguity. Fourthly, movies often belong to multiple genres, and in many cases, the assignment of genres can be subjective. This subjectivity adds an additional layer of complexity in accurately labeling movies within specific genres. Furthermore, genre labels do not capture the degree or intensity of a genre within a video. For instance, a movie like Jurassic Park can be classified as a science fiction and adventure film, but it also contains elements of horror and thriller. The genre labels alone fail to convey the nuanced blend of genres present in the movie. Moreover, movies within the same genre can still exhibit substantial differences. For example, consider two movies namely `Gladiator` and `Die Hard`, both categorized as action films. However, the flavor of action in these movies diverges significantly due to distinct contextual factors. Gladiator is an epic historical action film set in ancient Rome, showcasing thrilling battles and action sequences within the Colosseum. On the other hand, Die Hard centers around intense action scenes taking place in a modern skyscraper during a terrorist siege. 

\vspace{-5pt}
\section{Genre Spectrum}
We propose an alternative approach to examining the genres which we refer to as the \textit{Genre Spectrum}. Our hypothesis is that every title consists of a spectrum of genres and we transform the discrete genre labels data into a latent space where each dimension could be considered as an abstract concept/characteristic of a movie. Every genre will then manifest itself in this latent space  as a subspace defined by a range of combinations of all the latent dimensions. We hypothesize that the continuum nature of genre spectrum embeddings enhances its expressive power in comparison to the discrete genre labels. 
\begin{figure}[h]
    \centering
    \includegraphics[scale=0.25]{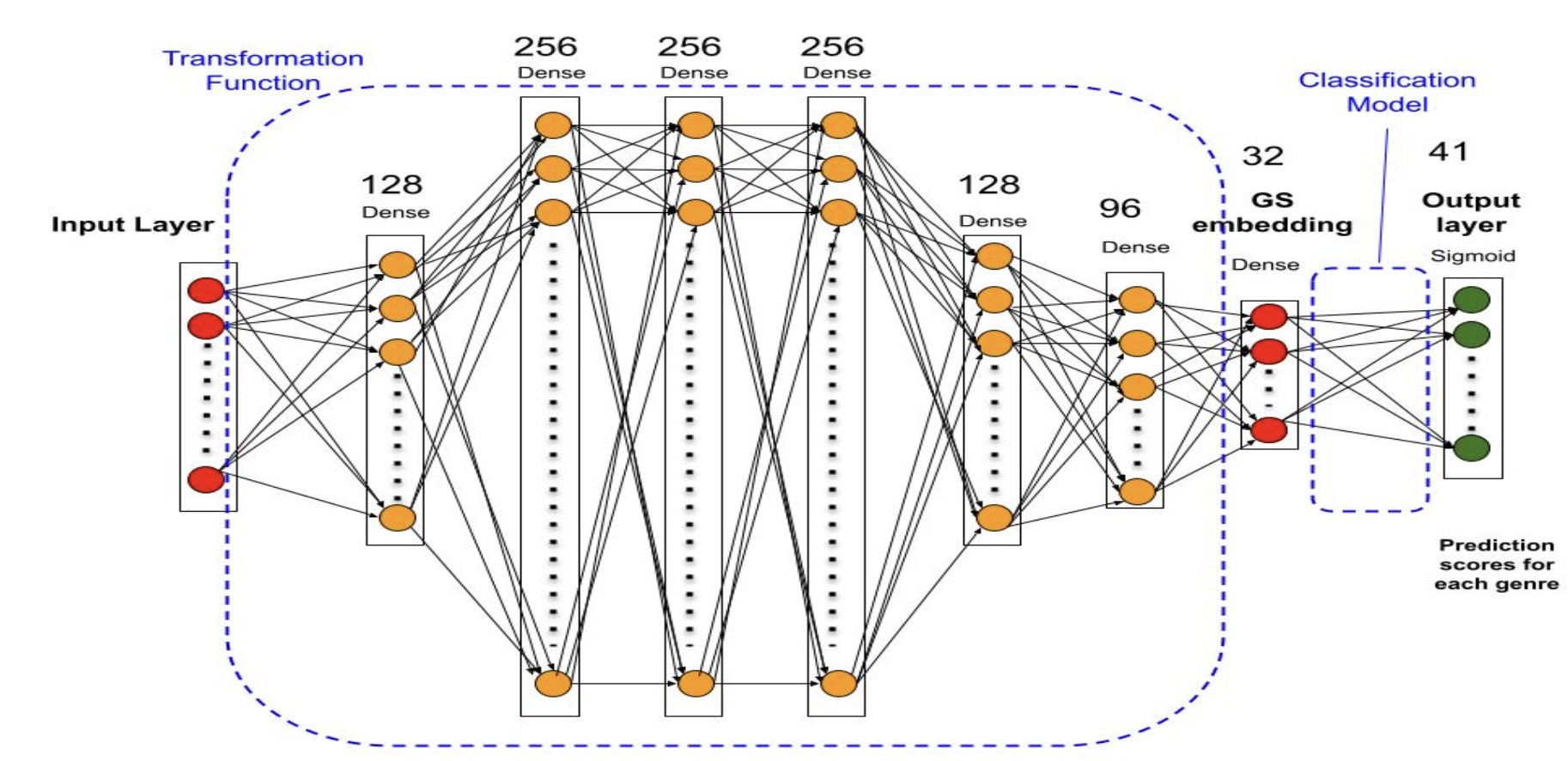}
    \caption{Neural net architecture for learning Genre Spectrum Embeddings}
    \label{fig:net_arch}
\end{figure}
\vspace{-10pt}
\subsection{Methodology}\label{sec:method}
We use neural network based supervised machine learning to learn genre-spectrum embeddings. The underlying intuition is that the textual metadata of movies (e.g. genre, language, year of release, plot synopsis and summary, ratings, Rotten Tomatoes scores, user reviews, box office information etc.) have rich information to classify a movie into one or more genre labels. We collect textual metadata of about 1.1M movies from various sources and apply language modeling techniques to learn textual embedding of every movie in a text-embedding space. We then formulate a multi-label classification problem that aims to predict the genre labels using learned textual embeddings as input features. In particular, we train a multi-layer feedforward dense neural network that ingests textual embeddings as inputs and emits the probabilities of every genre class as the output. The model is trained using cross-entropy loss averaged over all the genre classes. Both the components of the neural net, the textual-to-genre-spectrum transformer and the genre classifier are trained jointly on the multi-label cross-entropy loss function. Thus, once the model is trained, we simply obtain genre-spectrum embeddings, by doing a forward pass on the transformer component of the neural net, i.e. collect the output from the penultimate layer of the neural net (as shown in Figure ~\ref{fig:net_arch}).

\textbf{Data Augmentation:} To further improve the quality of embeddings particularly on less-popular movies that have poor quality of metadata, we applied data augmentation technique proposed in \cite{zhang2018mixup} on the training data. This technique randomly samples two training samples and take their random convex combination (on both features and labels) to generate a new synthetic data sample. We applied this technique (with small modifications to increase representation of rarer classes) and increased the training data by a factor of 10.  


\section{Experiments \& Results}
\subsection{Data and Setup}
We collected textual metadata and genre labels of about 1.1M movies and tv series from three sources of movie metadata, namely: IMDb, Rotten Tomatoes, and Gracenote. We used 60-20-20 split to generate training, validation, and test set. 

\begin{figure}
    \centering
    \includegraphics[scale=0.30]{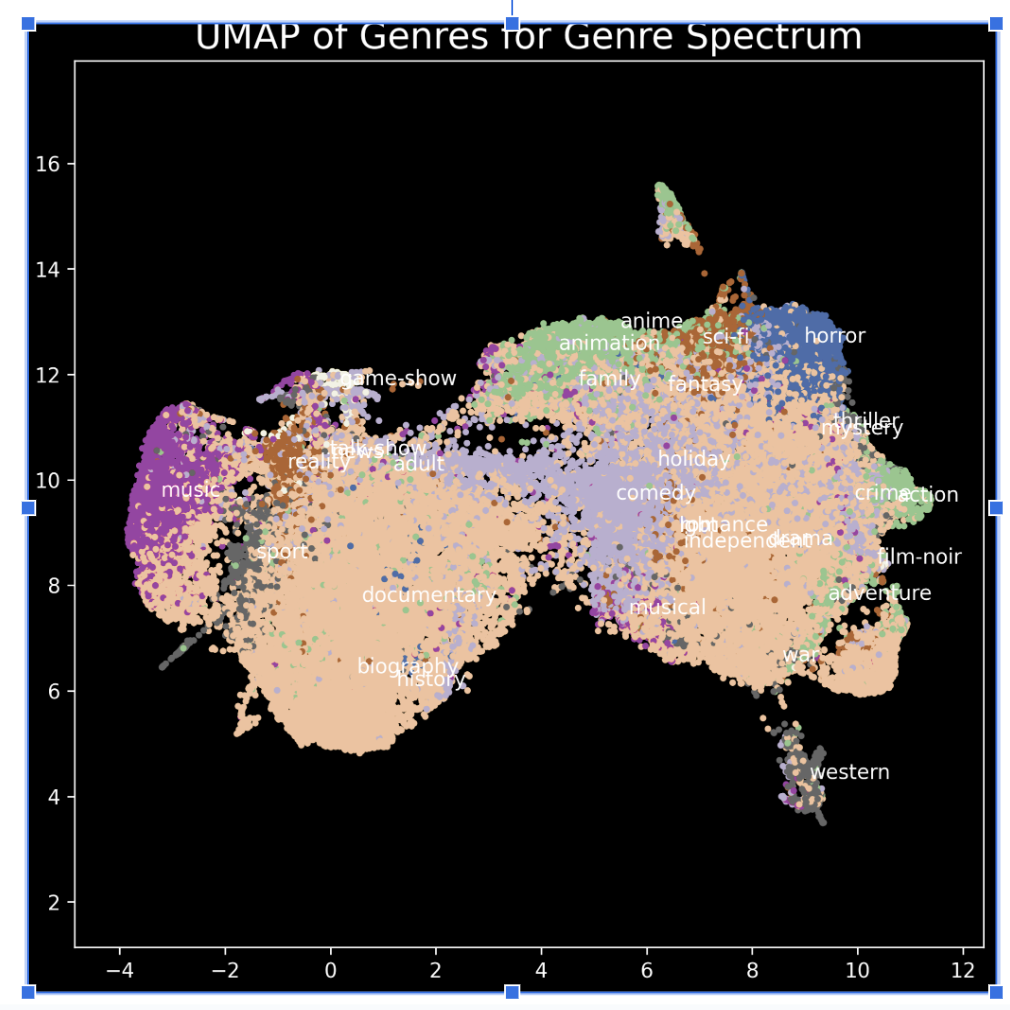}
    \caption{A 2-D plot of genre spectrum embeddings generated using Universal Manifold Approximation and Projection (UMAP) technique}
    \label{fig:genre_spectrum}
\end{figure}

\vspace{-7pt}
\begin{table*}[ht!]
\centering
\renewcommand{\arraystretch}{1.2} 
\caption{Comparison of latent feature spaces on genre similarity in top-100 neighborhoods}
\label{tab:offline_res}
\resizebox{0.8\textwidth}{!}{%
\vspace{-5pt}
\begin{tabular}{@{}p{2.3cm}|p{1.4cm}|p{1.4cm}|p{1.4cm}|p{1.4cm}|p{1.4cm}|p{1.4cm}|p{1.4cm}@{}}
\toprule
\textbf{Embeddings} & \multicolumn{7}{c}{\textbf{Top-100 nb genre-similarity score (\%) on popularity-based groups}} \\
\hline
& IMDb votes $\in [0, 10]$ & IMDb votes $\in [10, 100]$ & IMDb votes $\in [100, 1000]$ & IMDb votes $\in [10^3, 10^4]$ & IMDb votes $\in [10^4, 10^5]$ & IMDb votes $\in [10^5, 10^6]$ & IMDb votes $\in [10^6, 10^7]$ \\
\midrule
Doc2Vec (textual) & 65.60 & 68.96 & 76.66 & 84.07 & 88.08 & 88.54 & 92.30 \\
BERT (textual) & 40.27 & 44.64 & 53.07 & 60.89 & 65.26 & 64.74 & 72.44 \\
GPT-4 (textual) & 56.46 & 63.03 & 67.23 & 74.28 & 77.68 & 77.36 & 85.33 \\
GS on Doc2Vec & 78.50 & 83.28 & 89.84 & 94.43 & 96.48 & 96.45 & 97.89 \\
GS on BERT & 60 & 63.03 & 67.23 & 74.28 & 77.68 & 77.36 & 85.33 \\
GS on GPT-4 & 78.80 & 80.60 & 83.61 & 87.55 & 90.32 & 92.29 & 96.11 \\ 
GS-Augmented on Doc2Vec & 80.94 & 84.82 & 91.34 & 95.51 & 97.34 & 97.3 & 98.17 \\
\bottomrule
\end{tabular}%
}
\end{table*}


\subsection{Offline evaluation}
\begin{figure*}[h]
  \centering
  \includegraphics[width=\textwidth]{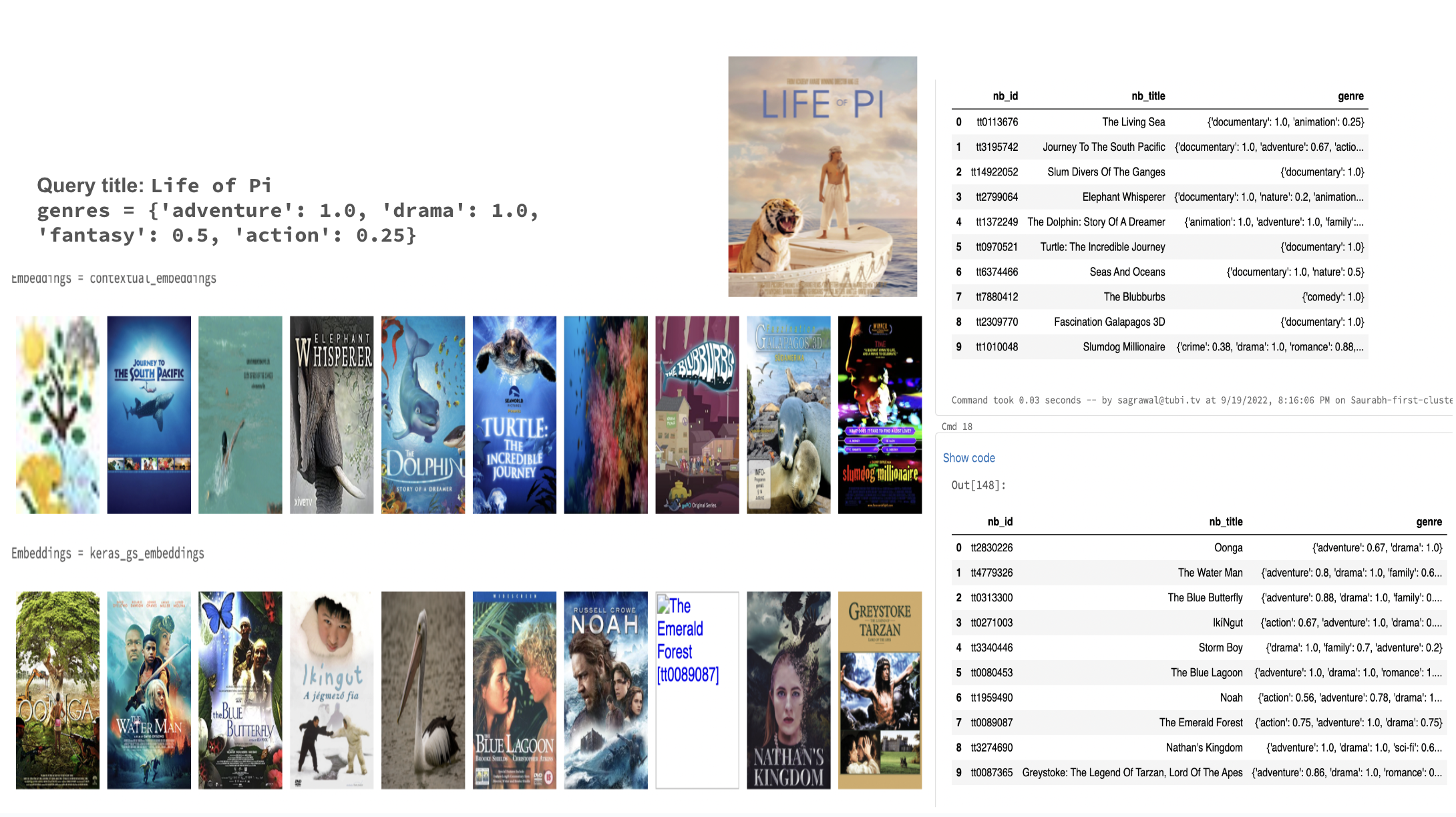}
  \caption{Comparison of top-10 neighbors of movie \textit{Life of Pi} in textual embeddings (Doc2Vec) space and the genre spectrum embeddings space (trained on Doc2Vec) in the top and bottom panel respectively.}
  \label{fig:lifeofpi}
\end{figure*}

For qualitative evaluation, we generated the 2-D plot of genre spectrum embeddings produced using UMAP(Uniform Manifold Approximation and Projection) technique. As can be seen, the genres appear to be cohesive colored clusters in the latent space. Next, we evaluate different variants of Genre Spectrum (GS) embeddings based on the genre similarity in the neighborhood of every movie. Specifically, for each movie, we compute genre similarity in top-$k$ neighborhood as a fraction of top $k$ nearest neighbors in genre-spectrum space that share one or more primary genres with the given movie. A primary genre is defined as the one that is assigned to a movie by majority of the labeling sources. To gain deeper insights into the relationship between the metric and popularity, the genre similarity score is calculated as an average across different subsets of movies, which are grouped based on their IMDB votes as shown in Table~\ref{tab:offline_res}. The first six rows in the table correspond to six variants of embeddings, three of them are textual embeddings generated using a variety of NLP models including Doc2Vec (96 dimensions)\cite{le2014distributed}, pretrained BERT model trained on web corpus \cite{wolf2019huggingface} (768 dimensions), and OpenAI GPT-4 (1536 dimensions) \cite{openai2023gpt4} the latest version of LLM released by OpenAI. The next three rows correspond to Genre Spectrum embeddings learnt using genre label supervision on each one of the aforementioned textual embeddings. The last row corresponds to another variant of \textit{GS on Doc2Vec} where we applied data augmentation step described in Section~\ref{sec:method}. We make several insightful observations from the table: i) All the variants of Genre Spectrum embeddings perform better than their corresponding textual embedding variants in all the popularity buckets, validating the effectiveness of our approach. In particular, the effectiveness of our proposed methodology also applies in the context of LLMs. ii) Further, it can be seen that the improvement in genre-similarity is higher on the lower popularity buckets. This could potentially be attributed to the fact that the quality of metadata (e.g. terse synopses, less tags) degrades on non-popular movies. Consequently, the textual embeddings tend to be more unreliable in classifying genres for such movies. However, the noise is considerably reduced in Genre Spectrum embeddings as they are trained using genre labels. iii) \textit{GS-Augmented on Doc2Vec} beats \textit{GS on Doc2Vec} consistently in genre similarity scores for all the popularity segments, justifying the utility of data augmentation step. Further in Figure~\ref{fig:lifeofpi}, we present an anecdotal example of a popular movie called \textit{Life of Pi}` to compare top-10 neighbors in textual and genre spectrum embedding spaces. In comparison to textual embedding space, neighbors in genre-spectrum latent space are much better aligned with the query movie on genre similarity.  

\subsection{Online evaluation}

To evaluate genre-spectrum embeddings in our online Tubi recommender system, we introduced a retrieval model in our production system. This model retrieves nearest neighbors of movies the user previously watched. Through an A/B test, we compared it to the control variant, which utilized binary genre labels. The test resulted in a statistically significant 0.6 \% improvement in our primary user-engagement metric, 'tvt-capped' (total view time capped at 4 hours per day). This improvement validates the effectiveness of genre spectrum embeddings in enhancing personalization and user engagement.

\section{Conclusion \& Future Work}
We presented a case study on various challenges in incorporating genre label information in movie recommendation systems and how to address those challenges by learning meaningful embeddings to capture genre label information in a video-recommendation problem setting. An evident expansion of our work involves broadening the scope of content metadata to encompass other manually annotated movie datasets that offer a more extensive range of tags. Nevertheless, a common hurdle with such datasets lies in their limited coverage. Given the powerful capabilities of LLMs, one of the potential future directions could be to apply LLMs on textual metadata and generate more specific annotations for every movie in the form of \textit{micro-genres}. Such micro-genres could then be used along with genre labels to learn more precise representation vectors of movies. Additionally, micro-genres could also be very useful in optimal organization of movie recommendations on user's home screen. In particular, movie recommendations on prominent Video on Demand (VOD) platforms such as Tubi, Netflix, and Amazon Prime are typically presented in a 2-D grid layout using a set of 'carousels.' Each carousel groups together movies with a common theme, such as genre, language, or year, as reflected in its title. Conventional methods often use limited themes (e.g., standard genres or 90's classics) for carousel generation, which might result in sub-optimal personalization of the home-grid. By incorporating LLM-generated micro-genres, we can enrich the pool of carousel themes, leading to more effective personalization. During the presentation, we will also share preliminary results from our explorations in this direction.

\section{Biographies}
\textbf{Saurabh Agrawal} is a Senior Machine Learning Engineer at Tubi since August 2022 where he leads deep learning projects for Search and Recommendation Systems at Tubi. Prior to Tubi, he completed his PhD in Computer Science from University of Minnesota before he worked at Amazon for more than three years as an Applied Scientist. 

\textbf{John Trenkle }is an experienced professional in AI/ML. John's work at Tubi includes significant contributions in Recommendation Systems, AdTech, Natural Language Processing (NLP), and Big Data management, showcasing his adaptable approach to the evolving field of machine learning.

\textbf{Jaya Kawale} is the VP of Engineering at Tubi leading all the machine learning efforts at Tubi. She did her PhD in Computer Science from the University of Minnesota and has published 15+ papers at top-tier machine learning conferences. Prior to Tubi, she has worked at Netflix, Adobe Research, Yahoo Research and Microsoft Research.




\bibliographystyle{ACM-Reference-Format}

\end{document}